Estimation of the leaf area density distribution of individual trees using high-resolution and multi-return airborne LiDAR data

Haruki Oshio[a,*], Takashi Asawa[b], Akira Hoyano[c] and Satoshi Miyasaka[d]

[a] Department of Environmental Science and Technology, Tokyo Institute of Technology, 4259-G5-2, Nagatsuta-cho, Midori-ku, Yokohama, Kanagawa, 226-8502, Japan, oshio.h.aa@m.titech.ac.jp

[b] Department of Environmental Science and Technology, Tokyo Institute of Technology, 4259-G5-2, Nagatsuta-cho, Midori-ku, Yokohama, Kanagawa, 226-8502, Japan, asawa.t.aa@m.titech.ac.jp

[c] The Open University of Japan, 2-11, Wakaba, Mihama-ku, Chiba, 261-8586, Japan, hoyano@ouj.ac.jp

[d] Nakanihon Air Service CO., LTD., 17-1, Toyobawakamiya, Toyoyama-cho, Nishikasugai, Aichi, 480-0202, Japan, miyasaka@nnk.co.jp

* Corresponding author. Tel.: +81 459245510; Fax: +81 459245553. E-mail address: oshio.h.aa@m.titech.ac.jp (Haruki Oshio). Postal address: 4259-G5-2, Nagatsuta-cho, Midori-ku, Yokohama, Kanagawa, 226-8502, Japan

ABSTRACT

In this paper we demonstrate a method for estimating the leaf area density (LAD) distribution of individual trees using high-resolution airborne LiDAR. This method calculates the LAD distribution from the contact frequency between the laser beams and leaves by tracing the laser beam paths. Multiple returns were used to capture the foliage distribution in the inner part of the crown. Each laser beam is traced from a location of the last return to the location of the first or intermediate return that is recorded immediately before the last return. We verified the estimation accuracy of the LAD distribution using terrestrial LiDAR data from single trees (*Zelkova serrata* and *Cinnamomum camphora*). The appropriate voxel size for representing the LAD distribution from the airborne LiDAR data was determined to be 1 m × 1 m × 0.5 m. The accuracy of the estimated LAD distribution for this voxel size was then examined while considering the number of airborne incident laser beams on the voxel ($N$) and the return type used. When only the first and single returns were used, the LAD was overestimated even for the voxels with large $N$. LAD was estimated as zero for most voxels with small $N$, although LAD was significantly overestimated for several voxels. We found that using the last and intermediate returns improved the LAD estimation accuracy even if $N$ was the same. The mean LAD estimation error was 0.25–0.3 $m^2/m^3$ for both species. Assigning different weights to the first and intermediate returns improved the accuracy slightly. Estimation error clearly corresponded to $N$, and $N$ of 8–11 could be a criterion for an accurate LAD estimation.

## 1. Introduction

Trees in urban spaces have a significant influence on urban environments through solar shading, transpiration, wind breaking, air purification and soundproofing. Knowledge of the three-dimensional structures of individual trees is important for their maintenance and for understanding their effects on urban environments. The leaf area density (LAD) distribution is a key index for characterizing the vertical and horizontal crown structures and is defined as the total one-sided leaf area per unit volume.

Various ground-based indirect methods for measuring LAD distribution have been developed. The point quadrat method (Wilson, 1960; Wilson, 1963) offers an accurate estimation when sufficient probes are inserted into the target canopy. Iio et al. (2011) developed a three-dimensional light transfer model based on this method, which can accurately estimate the photosynthetic photon flux density distribution in a crown. However, the point quadrat method is well known to be labor intensive. Another approach is the gap fraction method, which measures transmitted light under the target canopy and is often used to estimate foliage density. One implementation of this is the LAI-2000 (LI-COR) plant canopy analyzer (Welles & Norman, 1991) that has been used in numerous studies to obtain the leaf area index (LAI).

Techniques for estimating the influence of leaf clumping have also been studied (Chen et al., 1997; Ryu et al., 2010b), but three-dimensional foliage distribution remains difficult to measure. Recently, terrestrial light detection and ranging (LiDAR) has received much attention as a means of determining the canopy structure. Detailed tree models have been generated that reconstruct each shoot or leaf (Côté et al., 2009; Côté et al., 2011; Hosoi et al., 2011). High-resolution terrestrial LiDAR observation offers an accurate estimation of the LAD distribution (Hosoi & Omasa, 2006; Hosoi & Omasa, 2007). However, it is laborious to carry out such LiDAR scans for many trees.

Airborne small-footprint LiDAR can acquire three-dimensional information of many trees at a high spatial resolution in a short time. The following methods have been established for deriving the tree geometry from airborne LiDAR data: tree height (Omasa et al., 2000; Hyyppä et al., 2001; Næsset & Økland, 2002; Persson et al., 2002), crown base height (Holmgren & Persson, 2004; Popescu & Zhao, 2008), and crown shape and volume (Omasa et al., 2008; Hecht et al., 2008; Kato et al., 2009). LAI estimation has also been investigated using gap fractions (Solberg et al., 2009; Korhonen et al., 2011), regression models of LiDAR metrics (Riaño et al., 2004; Farid et al., 2008; Jensen et al., 2008; Zhao and Popescu, 2009) and the contact frequency between laser beams and foliage (Morsdorf et al., 2006). Wang et al. (2008) and Sasaki et al. (2012) also attempted a voxel-based reconstruction of the foliage distribution. However, these studies did not consider the LAD for each voxel. Solberg et al. (2010) used voxel-based gap fraction (laser penetration rate) distribution for simulating X-band interferometric height.

Methods for estimating LAD distribution using airborne LiDAR have been explored less thoroughly. Hosoi et al. (2011) developed a method for estimating LAD distribution by combining airborne and terrestrial LiDAR. They calculated the LAD based on the contact frequency between the laser beams and the leaves. The contact frequency was computed by tracing the path of the laser beams and counting the number of laser beams in each layer that trigger a return (laser beam interception) and those that do not (laser beam pass). An airborne LiDAR point cloud acquired by the first return mode was used, which yielded an underestimated LAD when only airborne data were used. Song et al. (2011) proposed a method for estimating plant area density (PAD) distribution involving the acquisition of airborne LiDAR data by employing a multi-return mode. Laser beams were traced from the points derived from only the first or single returns to avoid the multiple counting of individual laser beams.

This study aimed to develop a method for estimating the LAD distribution of individual trees using multi-return airborne LiDAR data. More specific objectives were to determine the appropriate voxel size for representing the LAD distribution by the airborne LiDAR data, and elucidate the estimation accuracy of LAD while considering the number of incident laser beams on the voxel and return type used.

## 2. Materials

### 2.1 Airborne LiDAR data

The study site was Hisaya-Odori Street in Nagoya, Japan. Hisaya-Odori Street is wide and lined with numerous broadleaved trees. A helicopter-based laser scanning system (Nakanihon Air Service) with an LMS-Q560 sensor (RIEGL) was employed for the airborne LiDAR observation. Fig. 1 (left) shows the flight track and Table 1 shows the data acquisition specifications. The flight altitude was 350 m, the footprint diameter was 0.18 m, and the distance between the consecutive footprint centers on the ground under the flight track was 0.2 m in the flight direction and 0.25 m in the scan direction. A multi-return mode was used; first, intermediate, last and single returns were obtained. The number of returns per laser shot that the sensor could obtain was unlimited. Airborne LiDAR data were acquired on September 6, 2010, during which there were leaf-on conditions in Japan.

We selected a Japanese Zelkova (*Zelkova serrata*) and a Camphor laurel (*Cinnamomum camphora*) for analysis. These trees are common roadside species in Japan. The crown of *Z. serrata* widens toward its upper part, in which the foliage is densely distributed. *C. camphora* has an oval crown and the foliage is distributed from the upper to the lower part of the crown. The height and foliage density of the selected trees were close to the averages for these species in the study site; each tree was selected from twenty trees of that species as discerned by analysis of the variation of tree height and laser beam penetration ratio ($P_{AL}$) using the airborne LiDAR data. $P_{AL}$ is given by $N_p/N_i$, where $N_p$ is the number of the laser beams penetrating the tree and $N_i$ is the number of the incident laser beams on the tree. For the *Z. serrata* and the *C. camphora*, the tree heights were 10.5 m and 15 m, the crown lengths were 7 m and 11 m, and the $P_{AL}$ were 0.37 and 0.32, respectively.

The location of the trees are shown in Fig. 1. The *Z. serrata* was an isolated tree and the *C. camphora* was a part of a canopy. The incident zenith angles of the laser beams from flight track 1 and flight track 2 were small for the *C. camphora* (mean zenith angle: 2.5°) and the *Z. serrata*, (mean zenith angle: 7.5°) respectively. The maximum number of returns per pulse was 4 for the *Z. serrata* and 5 for the *C. camphora*. The proportions of single, first, last, and intermediate returns to all returns from the crown were 0.35, 0.41, 0.11, and 0.13 for the *Z. serrate* and 0.28, 0.41, 0.18, and 0.13 for the *C. camphora*, respectively.

The foliage density varies within a tree crown. Therefore, even if only one tree is used, the relationship between the estimation accuracy of LAD for a voxel and the number of airborne laser beams incident on the voxel ($N$) can be examined while considering the variation in LAD. The relationship could be general knowledge to assess the estimation accuracy based on the obtained data. Trees in urban spaces differ in crown shape and foliage density because of pruning and health conditions. Ideally, the accuracy should be verified under all these conditions, but it is impractical to do so at one time. Therefore, we used trees with average structural characteristics to ensure broad variation in the LAD and $N$. The influence of the difference in foliage structure and foliage distribution on the estimation accuracy of LAD distribution was examined using different tree species.

*2.2 Field data acquisition and processing*

*2.2.1 Terrestrial LiDAR data*

Field measurements were carried out on September 2, 2010. We obtained the LAD distribution using terrestrial LiDAR to verify the results from the airborne LiDAR. The selected *Z.serrata* and *C. camphora* were measured from two scanning positions using a terrestrial laser scanner (VZ-400; RIEGL), as shown in Fig. 1. Table 2 shows the measurement specifications. The increments of the zenith and azimuth angles of the laser beam emission were 0.04° , yielding a point spacing of 7 mm on a vertical plane 10 m ahead.

*2.2.2 Principle of LAD calculation*

The LAD distribution of the trees was calculated using the terrestrial LiDAR data based on the method previously developed by Hosoi and Omasa (2006). This method is based on the point quadrat theory, given by the following equation,

$$LAD = \frac{1}{\Delta H} \cdot \frac{cos(\theta)}{G(\theta)} \sum_{k=1}^{N} \frac{n_i(k)}{n_i(k) + n_p(k)} \qquad (1)$$

where $\Delta H$ is the height of the voxel, $\theta$ is the incident laser beam zenith angle, $G(\theta)$ is the mean projection area of a unit leaf area on a plane perpendicular to the laser beam, $N$ is the number of layers in a voxel, $n_i(k)$ is the number of laser beam interceptions in the $k$th layer and $n_p(k)$ is the number of laser beam passes in the $k$th layer.

Hosoi and Omasa calculated $n_i(k)$ and $n_p(k)$ as follows. The LiDAR point cloud of the target canopy was divided into subvoxels of comparable size to the LiDAR spatial resolution; e.g., one of the voxels (1 m × 1 m × 1m) contained $10^6$ subvoxels (1 cm × 1 cm × 1cm). $n_i(k)$ was given by the number of subvoxels containing one or more returns in the $k$th layer. $n_p(k)$ was the number of subvoxels in the $k$th layer through which one or more laser beams passed. These voxels were detected by tracing the path of the laser beams.

### 2.2.3 LAD calculation

Integration of the point clouds acquired from multiple scanning positions causes LAD calculation error because the point clouds do not overlap each other owing to the movement of branches and leaves by wind (Asawa et al., 2014). Therefore, in this study the LAD distribution was calculated separately for each scanning position. The point clouds of the *Z. serrata* and *C. camphora* were divided into 1 m × 1 m × 0.5 m voxels and 5 mm × 5 mm × 5 mm subvoxels. The subvoxel size was decided from the terrestrial LiDAR spatial resolution and was the same size as in the work of Hosoi and Omasa (2007), who measured a *Z. serrata* canopy 9–24 m from the center of the canopy. $n_i(k)$, $n_p(k)$, and $\theta$ in Eq. (1) were calculated by tracing the path of the laser beams. $G(\theta)$ was calculated using the inclination angles of the leaves in each voxel. Terrestrial LiDAR points of foliage were selected from the upper, middle, and lower parts of the crown for each species. Thirty-five leaves were randomly selected from each foliage (i.e., a total 105 leaves were selected from each tree) and the frequency distribution of the leaf inclination angle with an interval of 10° was obtained. The average distribution acquired from the upper, middle, and lower parts was used because the difference in the distribution among heights was small. The LAD distribution of each scanning position was then generated.

The LAD for each voxel was derived from a scanning position having high terrestrial laser beam incident ratio (TLIR); i.e., the scanning position was voxel specific. The TLIR is given by $N_a/N_t$, where $N_a$ is the number of terrestrial incident laser beams on a voxel and $N_t$ is the number of terrestrial incident laser beams on a voxel when the laser beams are not intercepted by the objects between the voxel and the scanner. This type of approach has been used to evaluate laser beam interceptions before reaching the voxel (Durrieu et al., 2008; Béland et al.,

2011).

Voxels that were occupied by branches were manually detected and removed from analysis to

verify the accuracy of airborne LiDAR-derived LAD. More specifically, the point clouds of the

trees were manually classified into leaf points and wood points. Voxels where more than 90%

of the included points were leaf points were used for the verification. The reconstruction of

trees, including branch architecture, is left for future work. We were able to detect the terrestrial

LiDAR points of large branches because of their continuity and morphological characteristics.

The small branches (several millimeter in diameter) in the foliage were problematic, though,

because of occlusion effect. However, the influence of the small branches on the accuracy of terrestrial LiDAR-derived LAD is considered to be small. Hosoi and Omasa (2007) investigated the influence of non-photosynthetic tissues on the terrestrial LiDAR-based LAD estimation of *Z. serrata*. They showed that LAD was overestimated by 19% if LiDAR points of non-photosynthetic tissues were not eliminated. Hosoi et al. (2013) also estimated the volume of non-photosynthetic tissues of *Z. serrata*. They showed that the small branches (diameter less than 1 cm) occupied 24% of the total volume.

Voxels where there were sufficient number of terrestrial incident laser beams were selected to achieve an accurate verification. We used voxels with TLIR greater than 0.8. We have previously studied the accuracy of the LAD calculation using terrestrial LiDAR by direct measurement of leaf area (Asawa et al., 2014). For the same setup as in the present study (measurement distance, laser beam emission interval, voxel size), we confirmed that LAD was underestimated by 6% when TLIR was greater than 0.8. The terrestrial LiDAR-derived LAD error and the dispersion of this error increased when TLIR became less than 0.8. Therefore, in this study we corrected for the 6% LAD error in the voxels with TLIR greater than 0.8, and then the LAD was used for verifying the airborne LiDAR-derived LAD. After a single return, there are no further returns along that path direction behind the leaf because of an occlusion effect. However, this is expected to be small because there is relatively little leaf clumping on the spatial scale of a voxel.

Most of the selected voxels were located in the periphery and lower part of the crown. However,

we can discuss the relationship between $N$ and the LAD estimation accuracy because of the following reasons. For the voxels with large $N$, only a few were used for the verification because of the low TLIR in the upper part of the crown. However, the variation in the LAD estimation accuracy among these voxels is small because the airborne laser beams came into them uniformly. That is, the difference in the spacing of incident laser beams is small among voxels. For the voxels with small $N$, variation in the LAD estimation accuracy among voxels became large because of the non-uniform airborne incident laser beams. However, many of these voxels were used for the verification.

## 3. Methods

### 3.1 Analysis of the spatial resolution of the information derived from airborne LiDAR

Unlike the observation with the high laser beam density by terrestrial LiDAR, voxel size is important for airborne LiDAR-based estimation because the spatial resolution is in the order of 0.1 m. If there are too many 'unfilled voxels' from airborne LiDAR, the crown is not represented by the voxels. We define an unfilled voxel as one that contains no airborne LiDAR points but one or more terrestrial LiDAR points. Therefore, before estimating the LAD distribution, we examined the appropriate voxel size for the airborne LiDAR data to reduce the number of unfilled voxels. The point cloud acquired by flight tracks 1 and 2 was used for the *C. camphora* and the *Z. serrata*, respectively, in the following procedures. The airborne LiDAR point cloud of the trees were divided into voxels. Three different voxel sizes were used: 0.5 m × 0.5 m × 0.5 m, 0.75 m × 0.75 m × 0.5 m, and 1 m × 1 m × 0.5 m. Voxels containing one or more airborne LiDAR points were designated as being filled, i.e., volumes containing leaves. Voxels were also filled from terrestrial LiDAR data in the same way as airborne LiDAR. First, the filling ratio was calculated. The filling ratio is expressed as $N_{air \cap ter}/N_{ter}$, where $N_{air \cap ter}$ is the number of voxels filled by airborne LiDAR and terrestrial LiDAR, and $N_{ter}$ is the number of voxels filled by terrestrial LiDAR only.

Voxels with low LAD values are difficult to fill using airborne LiDAR. However, the influence on the reconstruction of the crown is small even if such low LAD voxels are not filled. The unfilled voxels were classified into three types: (i) the voxel contains insufficient leaves to trigger a return that the sensor can recognize (Voxel A); (ii) there are no incident laser beams on the leaves in the voxel, even if the voxel contains more leaves than Voxel A (Voxel B); and (iii) the voxel with TLIR less than 0.8 (Voxel C). If one of the voxels adjacent to an unfilled voxel is filled, the error is within one voxel. In particular, the influence of the unfilled voxel on the reconstruction of the crown is small when it is Voxel A. Therefore, Voxel A, Voxel B, and Voxel C having at least one of the adjacent voxels filled by airborne LiDAR were designated as Voxel A', Voxel B' and Voxel C'.

Fig. 2 shows the procedure for the classification. Terrestrial LiDAR-derived LAD was used to distinguish between Voxel A and Voxel B. First, voxels with a TLIR less than 0.8 were removed from the analysis (Voxel C; Fig. 2, Step 1). The threshold in Step 2 was calculated using the voxels where there were no objects between the voxel and the airborne LiDAR scanner. The maximum terrestrial LiDAR-derived LAD of the voxels containing no airborne LiDAR points was used for the threshold. If the LAD of an unfilled voxel was less than the threshold, the voxel was considered to contain insufficient leaves to trigger a return (Voxel A). Voxel A, Voxel B, and Voxel C were then classified based on whether the adjacent voxel was filled (Fig. 2, Step 3).

### 3.2 Estimating LAD distribution using airborne LiDAR data

### 3.2.1 Previous methods

Hosoi et al. (2011) estimated LAD distribution using airborne LiDAR and terrestrial LiDAR. The method was based on the same theory as Hosoi and Omasa (2006); see Section 2.2. $n_i(k)$ and $n_p(k)$ in Eq. (1) were detected by tracing the path of the laser beams. An airborne LiDAR point cloud acquired by the first return mode was used, which yielded an underestimated LAD when only airborne data were used. In their method, if inaccurate trajectory data for pulses (e.g., Song et al., 2011) are used, there exists the possibility of a ray failing to hit a subvoxel containing an airborne LiDAR return.

Song et al. (2011) estimated the PAD distribution using the same principle described by Eq. (1). They traced each laser beam between the location of the first or single return and the airborne LiDAR scanner using flight track information. Laser beam interceptions and passes were counted in each voxel. The last and intermediate returns were excluded to avoid multiple counting of individual laser beams. It was assumed that the flight track was perpendicular to the scan line.

Reconstruction of the laser beam interceptions and passes, including the last and intermediate returns, may be accomplished by using the trajectory data for all pulses. The trajectory data are acquired from the smoothed best estimated trajectory, which contains the flight track and orientation of the platform (e.g., Korpela et al., 2012). Our method performs the reconstruction solely from return locations.

*3.2.2 Proposed method*

Fig. 3 shows a schematic of the method proposed in this study. The point clouds of the *Z. serrata* and *C. camphora* were divided into voxels representing the LAD distribution. Each voxel was divided into several layers, and $n_i(k)$ and $n_p(k)$ from Eq. (1) were calculated in each layer. The reason for introducing the layers within a voxel is as follows. When a single return or a last return is triggered, there are no further returns along the laser beam path direction. Even for a first return or an intermediate return, there are no further returns in the voxel because the voxel height is of the same order as the range resolution of the airborne LiDAR. Therefore, leaves behind a return are not counted when the voxel is the smallest spatial unit, yielding an underestimated LAD. We assumed that the leaf distribution behind a return is approximated by the leaf distribution in the area that was illuminated by the laser beams in the same layer. We selected a layer thickness of 0.1 m as per the previous study (Hosoi et al., 2011). The height of the voxel was 0.5 m (§3.1), yielding five layers in the voxel.

The value of $n_i(k)$ was given by the number of points in the layer. $n_p(k)$ was calculated from laser beam tracing as follows: for the last or intermediate returns, each laser beam was traced from the return location to the location of the first or intermediate return that was recorded just before the last or intermediate return. This tracing is possible because locations of multiple returns to one pulse are recorded on data file in succession, and the return location that was recorded before or after the target return can be detected. For the first and single returns, each laser beam was traced from the return location to the position of the scanner. The direction of the tracing was determined as follows. For the first returns, the direction to the last or intermediate return that was recorded just after the first return was used. For the single returns, the average direction of the tracing for the first returns was used. The last and intermediate returns can easily be included in the calculation without missing traces. No information is needed regarding the flight track or the position and direction of the laser beam emission. $G(\theta)$ was calculated using the frequency distribution of leaf inclination angle obtained by terrestrial LiDAR data of the trees (§2.2).

Hosoi et al. (2011) and Song et al. (2011) regarded a first return as completely intercepting the laser beam (i.e., 1 was added to $n_i(k)$ in Eq. (1) ). However, the laser beam footprint of a first return is not fully occupied by the leaves. Therefore, we assigned 0.6 laser beam interceptions to the points derived from the first returns (0.6 was added to $n_i(k)$ in Eq. (1)) for both *Z. serrata* and *C. camphora*. The value 0.6 was the ratio of the average reflection intensity of the first returns to that of the single returns. The intermediate returns were also assigned 0.6 interceptions.

The difference between return location (footprint center) and location of the target is less than footprint radius. We assume that the influence of this difference on the estimation accuracy is negligible. The relationship between the LAD and the contact frequency between laser beams and leaves is theoretically formulated as Eq. (1) if the laser beam spacing and laser beam divergence are much smaller than the leaf and if all regions are illuminated by laser beams. We assume that this contact frequency can be approximated by processing the airborne LiDAR data, as depicted in Fig. 3, especially for the voxels where there are sufficient incident laser beams. Terrestrial LiDAR data were used to obtain the leaf angle distribution in the present study. For an operational application of the method, if it is difficult to conduct terrestrial laser scanning then easier methods are available, such as digital camera-based methods (e.g., Ryu et al., 2010a), and methods based on standard distributions (e.g., spherical, erectophile, plagiophile, and planophile distributions (de Wit, 1965), the beta distribution (Goel & Strebel, 1984), and the elliptical distribution (Kuusk, 1995)). A database of the best approximation model and parameters for each tree species (e.g., Pisek et al., 2013) is needed. Information on tree species cen be obtained by hyperspectral imaging or LiDAR data (e.g., Alonzo, 2014) and is collected by municipalities in urban areas in Japan.

## 4. Results and Discussion

*4.1 Comparison between airborne and terrestrial LiDAR-derived voxel distributions*

Fig. 4 shows the vertical distribution of the filling ratio according to height from crown base.

There were many incident laser beams on the voxels in the upper part of the crown. However,

the filling ratio was only 40–70% for a 0.5 m × 0.5 m × 0.5 m voxel. The filling ratio decreased to less than 40 % in the middle to lower part of the crown. The difference in the filling ratios between 0.75 m × 0.75 m × 0.5 m and 1 m × 1 m × 0.5 m voxels was small in the area near the treetop. The difference increased, though, toward the lower part of the crown, exceeding 15%.

The filling ratio in the entire crown area is also displayed in Fig. 4. The filling accuracy was relatively low for 0.5 m × 0.5 m × 0.5 m and 0.75 m × 0.75 m × 0. 5 m voxels; we chose a voxel size of 1 m × 1 m × 0.5 m. The possible reasons for the unfilled voxels (i.e., voxels not being filled by airborne LiDAR) are analyzed in the next section.

*4.2 Characteristics of unfilled voxels*

The thresholds of the terrestrial LiDAR-derived LAD for classifying the unfilled voxels into

Voxel A and Voxel B (Fig. 2, Step 2) were 0.06 m$^2$/m$^3$ for the *Z.serrata* and 0.10 m$^2$/m$^3$ for the

*C. camphora*. The threshold was the maximum terrestrial LiDAR-derived LAD of the voxels

containing no airborne LiDAR points (§3.1). Fig. 5 shows the breakdown of unfilled voxels according to height. First, we describe the result of *Z. serrata*. There were more Voxel A' than Voxel B' in the upper half of the crown. There were some Voxel C' in this area. However, the main cause of these unfilled voxels was thought to be an insufficient number of leaves necessary to trigger a return corresponding to Voxel A'. In this area, all unfilled voxels had at least one adjacent voxel that was filled by the airborne LiDAR. Therefore, the influence of the unfilled voxels on the reconstruction of the foliage distribution was minimal. Although the number of Voxel B' and Voxel B increased toward the lower part of the crown, the proportion of voxels B' (i.e., at least one adjacent voxel was filled) remained high. The proportion of Voxel C' was higher for *C. camphora* than for *Z. serrata* because it had a larger crown hence lower TLIR. There were more Voxel A' than Voxel B' in the upper and middle part of the crown in both tree types.

The ratio of filled voxels and Voxel A' to all voxels except Voxel C and Voxel C' for the entire crown area was 85.6% for the *Z. serrata* and 89.2% for the *C. camphora*. In the upper and middle parts of the crown, which strongly influence its light absorption and transmission, this ratio was 95.2% in the 2–7 m height range (from the crown base) for *Z. serrata* and 97.8% in the 3–11 m height range for *C. camphora*. We conclude that the LAD distribution can be represented by the 1 m × 1 m × 0.5 m voxels.

*4.3 LAD distribution derived from airborne LiDAR*

Fig. 6 shows the leaf inclination angle distribution derived from the terrestrial LiDAR data and the variation of G($\theta$) according to the incident zenith angle of laser beams. The angle distributions of both trees are similar to the plagiophile distribution (de Wit, 1965). G($\theta$)$_{\text{plagio}}$/G($\theta$)$_{\text{meas}}$ = 0.95 for the *Z. serrata* and 1.0 for the *C. camphora* at the mean incident zenith angle of the laser beams from the flight track, where G($\theta$)$_{\text{plagio}}$ and G($\theta$)$_{\text{meas}}$ are G($\theta$) derived from the plagiophile distribution and the terrestrial LiDAR-based angle distribution, respectively. Therefore, the theoretical distribution has little influence on the estimation accuracy of airborne LiDAR-derived LAD when using the plagiophile distribution.

Several theoretical distributions have been proposed (§3.2.2). Pisek et al. (2013) reported that the planophile distribution was suited to *Z. serrata*; G($\theta$)$_{\text{plano}}$/G($\theta$)$_{\text{meas}}$ = 1.19 at the mean incident zenith angle of the laser beams from the flight track, where G($\theta$)$_{\text{plano}}$ is G($\theta$) derived from planophile distribution. Thus the estimated LAD with the planophile distribution is 0.84 (=1/1.19) times the LAD estimated from the terrestrial LiDAR-derived angle distribution. Researchers have measured the leaf inclination angle distribution of the roadside trees in Japanese urban areas (Shimojo et al., 2003; Yoshida et al., 2006). The angle distributions of *Z. serrata* and *C. camphora* were similar to those in this study. Therefore, a database of the best suited theoretical distribution is needed for each country or region.

Fig. 7 shows scatter plots of the airborne and terrestrial LiDAR-derived LAD. Fig. 7(1) shows the result when only the first and single returns were used, Fig. 7(2) is the result of all returns, and the effect of assigning 0.6 laser beam interception to the first and intermediate returns is shown in Fig 7.(3). The six different marker types indicate the quantity of incident laser beams on a 1 m × 1 m × 0.5 m voxel.

Hosoi et al. (2011) proposed a laser beam coverage index $\Omega$ that represents the area covered by the incident laser beams in each layer, taking into account the laser beam attenuation and the LiDAR's laser beam settings. The index $\Omega$ was a good measure of the LAD estimation error for both terrestrial and airborne LiDAR. However, the area covered by the laser beam footprint is difficult to determine for the last and intermediate returns. Furthermore, two laser beams with a footprint area of 1 are expected to offer better LAD estimation than one laser beam with a footprint area of 2. Thus, the number of airborne incident laser beams on each voxel is important. On this basis, we examined the relationship between $N$ and the LAD estimation accuracy. Fig. 8 and Fig. 9 show the relationships between $\Delta LAD$ (= |Airborne LiDAR-derived LAD – Terrestrial LiDAR-derived LAD|) and $N$, and between $R^2$ and $N$, respectively. The $\Delta LAD$ were averaged for each $N$ class. Fig. 10 shows the histogram of $N$ for the voxels where the terrestrial LiDAR-derived LAD was greater than zero (leaf voxels). Mean $\Delta LAD$ for all leaf voxels was calculated from Fig. 8 and Fig. 10 and is shown in Fig. 8.

### 4.3.1 Results for Z. serrata

When only the first and single returns were used (Fig. 7(1)), the airborne LiDAR-derived LAD was strongly correlated with the terrestrial LiDAR-derived LAD in the voxels where $N$ was large ($R^2$ exceeds 0.8 for $N$ greater than 12). Nevertheless, LAD was overestimated for most voxels (see Fig. 7(1)). There were many voxels where $N$ was 0–3 and the estimated LAD was 0 because these voxels contained no airborne LiDAR points derived from the first or single returns. On the other hand, LAD was significantly overestimated for several voxels having small $N$ because most of the incident laser beams hit leaves and triggered return even if the leaf area in the voxel was small. This often occurs when the number of incident laser beams is small. As a result, the mean $\mathit{\Delta LAD}$ was 0.48 m²/m³, 0.96 m²/m³, and 0.62 m²/m³ for the voxels where $N$ was 8–11, 4–7, and 0–3, respectively. In more than half of the leaf voxels $N$ was 0–7 when only the first and single returns were used (Fig. 10), resulting in a mean $\mathit{\Delta LAD}$ for all leaf voxels of 0.50 m²/m³.

Next, we examine the influence of using the last and intermediate returns (Fig. 7 (2)). $\mathit{\Delta LAD}$ was improved even if the number of incident laser beams on the voxel was the same (Fig. 8 (1) and (2)). This is because the number of laser beam passes was underestimated when only the first and single returns were used, as the laser beams between the first and last or intermediate returns were not reconstructed. More specifically, after a first return was recorded in a layer, 1 was added to $n_p(k)$ (Eq. (1)) in the subsequent layers in the laser beam path when the last and intermediate returns were included in the calculation. Although substantial overestimations occurred for voxels where $N$ was small, the number of such voxels dramatically decreased for voxels with $N$ of 8–11, 4–7, and 0–3. However, $\mathit{\Delta LAD}$ did not decrease for the voxels with $N$ of 0–3 because LAD was estimated as zero for most voxels when only the first and single returns were used, thus $\Delta LAD$ was small for the voxels in which the actual LAD was close to zero.

From Fig. 8(2), it is clear that $\Delta LAD$ corresponded to $N$ when all returns were used. $\Delta LAD$ decreased to 0.2 $m^2/m^3$ at an $N$ of 8–11 and varied slightly at larger $N$ values. More than 80% of the leaf voxels had $N$ greater than 4–7, and the mean $\Delta LAD$ for all leaf voxels decreased to 0.25 $m^2/m^3$. When the first and intermediate returns were counted as 0.6 laser beam interception, the most accurate estimation was achieved, although the decrease in the mean $\Delta LAD$ was small; the mean $\Delta LAD$ for all leaf voxels was 0.22 $m^2/m^3$.

*4.3.2 Results for C. camphora*

The proportion of the voxels where LAD was estimated as zero was higher for *C. camphora* than for *Z. serrata*. This is because the *C. camphora* had a greater crown length, meaning that there were fewer incident terrestrial laser beams in the upper part of the crown. Consequently, most of the used voxels came from the lower part of the crown. Mean $\Delta LAD$ for all leaf voxels was 0.43 $m^2/m^3$ when only the first and single returns were used. If all returns were used, although there was no plateau in the relationship between $\Delta LAD$ and $N$ (Fig. 8(2)), the order of the $\Delta LAD$ was the same as for the *Z. serrata*. Another similarity with the *Z. serrata* is that $\Delta LAD$ significantly decreased for the voxels with $N$ 4–7 and 8–11. Approximately 70% of the leaf voxels had $N$ greater than 4–7 (Fig. 10). The mean $\Delta LAD$ for all leaf voxels decreased to 0.29 $m^2/m^3$. When the first and intermediate returns were counted as 0.6 laser beam interceptions, the mean $\mathit{\Delta LAD}$ was a minimum of 0.26 m$^2$/m$^3$.

### 4.3.3 Consideration of the influence of foliage structure and crown structure

The relationship between the airborne and the terrestrial LiDAR-derived LAD only varied slightly for the *Z. serrata* and *C. camphora* in the voxels having large $N$. This means that the difference in foliage structure of *Z. serrata* and *C. camphora* had little influence on the airborne LiDAR-derived LAD. This is also supported by the fact that the mean $\mathit{\Delta LAD}$ for all leaf voxels of *C. camphora* decreased from 0.29 m$^2$/m$^3$ to 0.26 m$^2$/m$^3$ in the same way as *Z. serrata* when the first and intermediate returns were counted as 0.6 laser beam interceptions.

When a single laser beam interception was assigned to all returns, the tendency for LAD to be overestimated can be seen for voxels having large $N$ for *Z. serrata* and *C. camphora*, even if the last and intermediate returns were included in the calculation (Fig. 7 (2)). Relative error was large for voxels with small LAD. In the present experiments, assigning 0.6 laser beam interceptions to the first and intermediate returns improved the estimation accuracy. However, it is likely that the appropriate interception values vary according to the foliage structure. In other words, there is a possibility that a value is too small for dense foliage and too large for sparse foliage. Foliage density varies with species, growing conditions, and pruning, which affects reflection intensity. An LAD estimation taking into account the reflection intensity is left for future work.

For the voxels where $N$ was small, the percentage of voxels with overestimated LAD was greater for the *C. camphora* than the *Z. serrata*. This may be attributed to their different crown structures. When most of the incident laser beams on a voxel hit leaves and trigger returns, even if the leaf area in the voxel is small, the LAD is significantly overestimated. Although the last and intermediate returns were used, significant overestimation occurred when $N$ decreased to 0–3. On the other hand, the number of voxels where LAD was underestimated increased with decrease in number of incident laser beams on a voxel. This is because the probability that a laser beam hitting a leaf for the lower part of the crown is lower than for the upper part since laser beams travel through gaps. These overestimations and underestimations are caused by the non-random foliage distribution. The balance of underestimation and overestimation is expected to vary according to the foliage density and foliage distribution. Additional work is required to assess the balance for trees with different structures.

In this study, leaf voxels were detected by terrestrial LiDAR and were used for the verification of the airborne LiDAR-derived LAD. In an application of the proposed method, it is impractical to distinguish LAD and PAD by terrestrial LiDAR for large numbers of trees. For deciduous trees, one may be able to distinguish LAD and PAD by using leaf-on and leaf-off airborne LiDAR data. For evergreen trees, methods combining LiDAR data and tree architecture models (e.g., Côté, J.-F. (2009), Côté, J.-F. (2011)) have the potential to determine the influence of stems and branches. To develop these methods, more work on the relationship between LAD (PAD) distribution derived by the proposed method and the physical structure of trees is required.

## 5. Conclusions

We proposed an airborne LiDAR-based method for estimating the LAD distribution of individual trees. We calculated the contact frequency between the laser beams and leaves, while considering the multiple returns to a single incident pulse. We verified the proposed method using *Z. serrata* and *C. camphora* trees having average structural characteristics. The LAD distributions of the *Z. serrata* and the *C. camphora* derived from terrestrial LiDAR were used to verify the accuracy of airborne LiDAR-derived LAD distribution. The appropriate voxel size for the airborne LiDAR data was determined. There were many 0.5 m × 0.5 m × 0.5 m voxels that contained no airborne LiDAR points even if the voxels contained the terrestrial LiDAR points. A total of 60–70% of the leaf voxels were filled by airborne LiDAR when the voxel size was 1 m × 1 m × 0.5 m. Most of the unfilled voxels contained only a few leaves that were insufficient to trigger a return. The influence of these unfilled voxels on the reconstruction of the LAD distribution was small. We subsequently selected this voxel size to represent the LAD distribution.

When only the first and single returns were used, LAD was overestimated, even for the voxels where there were many airborne incident laser beams, although the estimated LAD strongly correlated to the terrestrial LiDAR-derived LAD. On the other hand, LAD was estimated to be zero for most voxels because the first and single returns were distributed only for the upper part of the crown. For several voxels where there were few airborne incident laser beams, LAD was significantly overestimated. Utilization of the last and intermediate returns decreased the estimation error even if $N$ was the same, especially for voxels with $N$ of 4–7 and 8–11. The percentage of voxels with larger $N$ also increased, leading to improved estimation accuracy for the entire crown area. Assigning 0.6 laser beam interceptions to the first and intermediate returns decreased the LAD estimation error slightly. The number of incident laser beams on each voxel was strongly correlated with the estimation error. More than eight incident laser beams on a 1 m × 1 m × 0.5 m voxel offered LAD estimation with an error of approximately 0.2 $m^2/m^3$. It is likely that this number of incident laser beams is a criterion for an accurate LAD estimation of *Z. serrata* and *C. camphora*.

In this paper we presented a fundamental method. To understand the robustness of this method tests across different foliage densities and crown structures are needed. To apply this method to typical airborne LiDAR campaigns, species-specific, regionally tuned databases of leaf inclination and methods for distinguishing LAD and PAD will be needed.

**Acknowledgments**

This work was supported in part by a Grant-in-Aid for JSPS Fellows. We would like to express gratitude to RIEGL Japan for their assistance with the terrestrial LiDAR observations.

5/W12, 209-212.

**List of Figure Captions**

Fig. 1. Airborne LiDAR observation flight tracks, location of the trees used in analysis, and terrestrial LiDAR scanning positions.

Fig. 2. Procedure of classification of unfilled voxel (voxel containing no airborne LiDAR points but having LAD > 0).

Fig. 3. Schematic of the proposed method.

Fig. 4. Variation of the filling ratio according to height. Filling ratio is ratio of voxels containing one or more airborne LiDAR points to voxels with LAD >0. The variation is plotted for each voxel size. Top: *Zelkova serrata*, Bottom: *Cinnamomum camphora*.

Fig. 5. Breakdown of voxels with LAD > 0. Voxel A: the voxel containing insufficient leaves to trigger a return that the sensor can recognize; Voxel B: there are no incident laser beams on the leaves in the voxel, even if the voxel contains more leaves than Voxel A; Voxel C: the voxel the LAD of which is not accurately calculated by terrestrial LiDAR. Single quote means that at least one adjacent voxel was a filled voxel. Proportion for each height and entire crown are displayed. Top: *Zelkova serrata*, Bottom: *Cinnamomum camphora*.

Fig.6. Leaf inclination angle distribution of *Zelkova serrata* (left) and *Cinnamomum camphora* (middle) derived from terrestrial LiDAR data. In the rightmost plot is G($\theta$), which is derived

from leaf inclination angle distribution acquired by terrestrial LiDAR and a theoretical distribution; black line, *Z. serrata*; black broken line, *C. camphora*; gray line, plagiophile distribution; gray broken line, planophile distribution; gray dotted line, spherical distribution; and the vertical line and vertical broken line indicate the mean incident zenith angle of the laser beams from the flight track for *Z. serrata* and *C. camphora*, respectively.

Fig. 7. Scatter plot of airborne LiDAR-derived LAD and terrestrial LiDAR-derived LAD. Results are plotted for each number of incident laser beams on a voxel ($N$). The symbol indicates $N$. Black ring: >20, Gray ring:16–19, Black cross:12–15, Gray cross:8–11, Black circle:4–7, and Gray circle:0–3. Leftmost: Results when only first and single returns were used; Middle: Results when all returns were used; Rightmost: Results when 0.6 laser beam interception was assigned to first and intermediate returns. Top: *Zelkova serrata*, Bottom: *Cinnamomum. camphora*.

Fig. 8. Variation of absolute difference between airborne LiDAR-derived LAD and terrestrial LiDAR-derived LAD ($\Delta LAD$) according to number of incident laser beams on a voxel ($N$). $\Delta LAD$ was averaged for each $N$ class. Mean $\Delta LAD$ shown at above right of each graph is the mean $\Delta LAD$ for all leaf voxels. This was calculated using the histogram of $N$ for the leaf voxels. Top: *Zelkova. serrata*, Bottom: *Cinnamomum camphora*.

Fig. 9. Variation of $R^2$ according to number of incident laser beams on a voxel ($N$). Top: *Zelkova. serrata*, Bottom: *Cinnamomum camphora*.

Fig. 10. Proportion of number of incident laser beams on a voxel ($N$) for all leaf voxel. Gray: Only first and single returns were used, Black: All returns were used. Top: *Zelkova. serrata*, Bottom: *Cinnamomum camphora*.

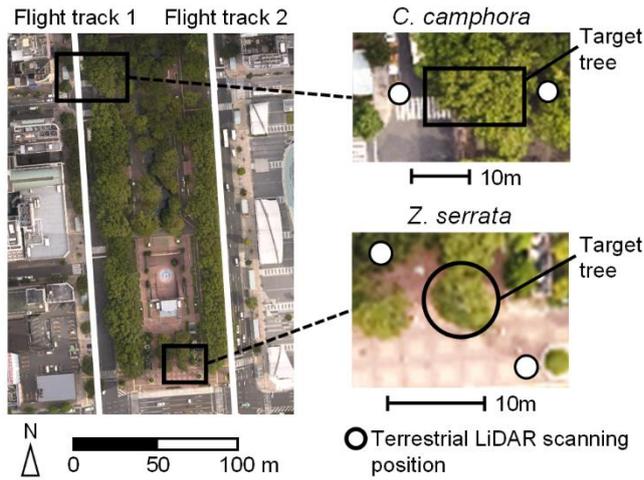

Fig. 1. Airborne LiDAR observation flight tracks, location of the trees used in analysis, and terrestrial LiDAR scanning positions.

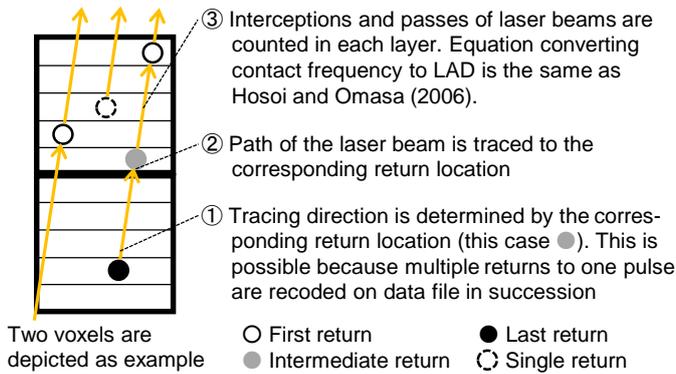

Fig. 3. Schematic of the proposed method.

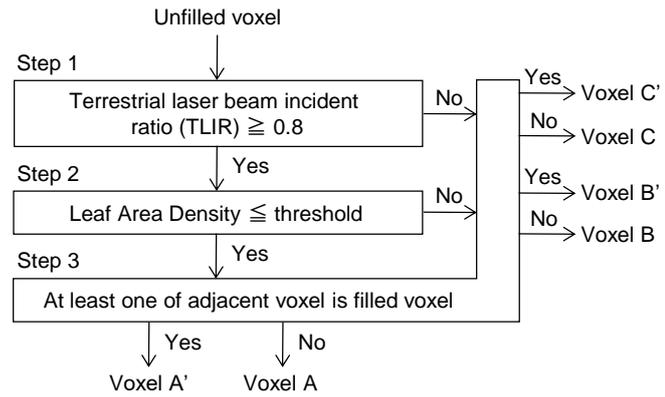

Fig. 2. Procedure of classification of unfilled voxel (voxel containing no airborne LiDAR points but having LAD > 0).

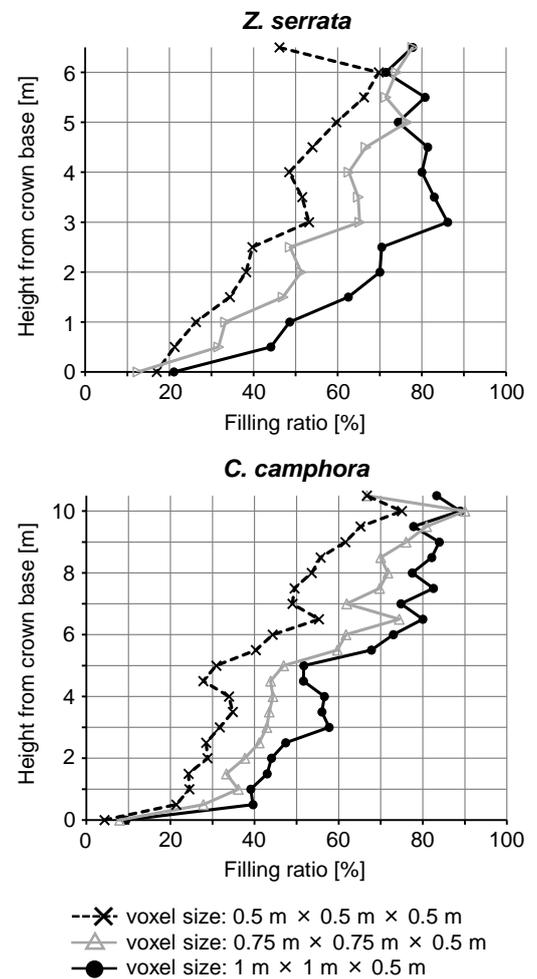

voxel size: 0.5 m × 0.5 m × 0.5 m
voxel size: 0.75 m × 0.75 m × 0.5 m
voxel size: 1 m × 1 m × 0.5 m

Fig. 4. Variation of the filling ratio according to height. Filling ratio is ratio of voxels containing one or more airborne LiDAR points to voxels with LAD >0. The variation is plotted for each voxel size. Top: *Zelkova serrata*, Bottom: *Cinnamomum camphora*.

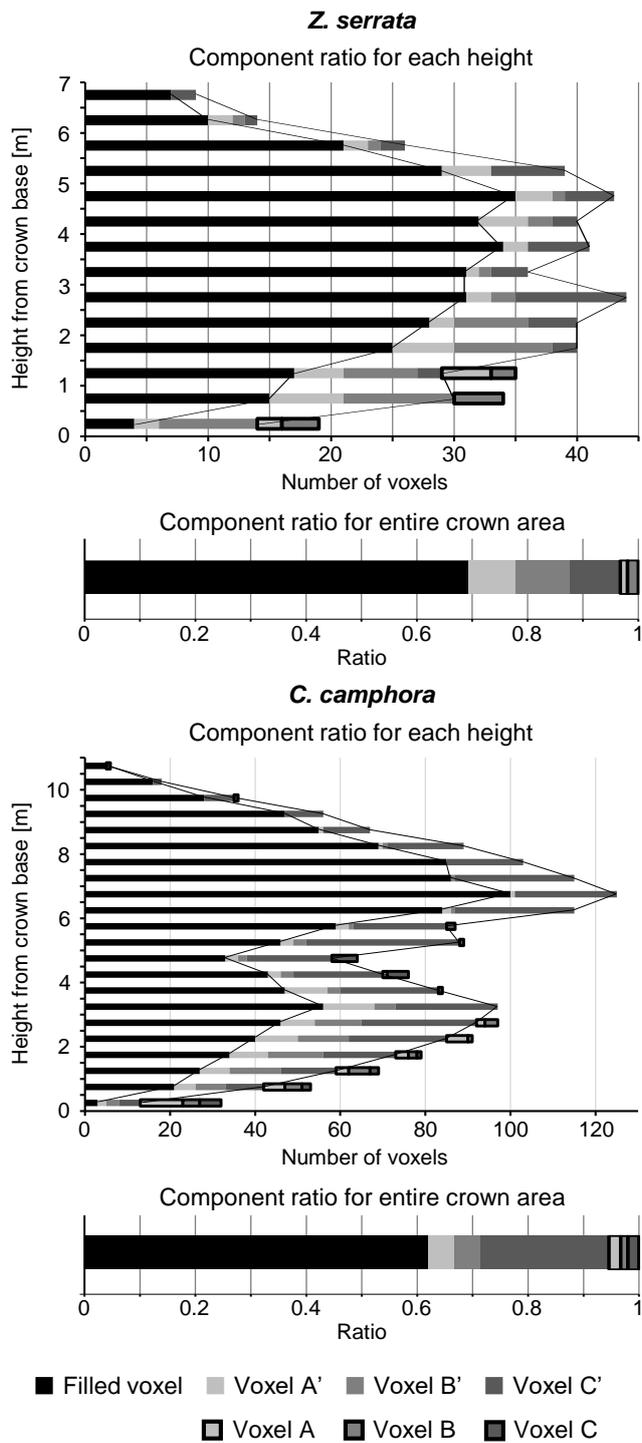

Fig. 5. Breakdown of voxels with LAD > 0. Voxel A: the voxel containing insufficient leaves to trigger a return that the sensor can recognize; Voxel B: there are no incident laser beams on the leaves in the voxel, even if the voxel contains more leaves than Voxel A; Voxel C: the voxel the LAD of which is not accurately calculated by terrestrial LiDAR. Single quote means that at least one adjacent voxel was a filled voxel. Proportion for each height and entire crown are displayed. Top: *Zelkova serrata*, Bottom: *Cinnamomum camphora*.

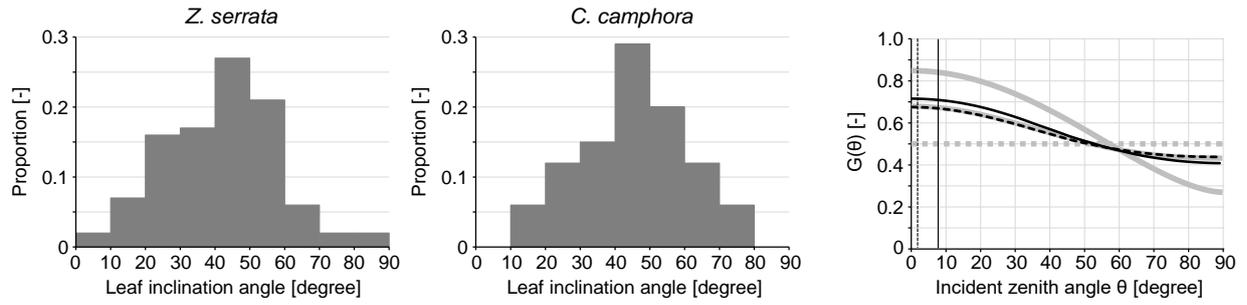

Fig. 6. Leaf inclination angle distribution of *Zelkova serrata* (left) and *Cinnamomum camphora* (middle) derived from terrestrial LiDAR data. In the rightmost plot is G($\theta$), which is derived from leaf inclination angle distribution acquired by terrestrial LiDAR and a theoretical distribution; black line, *Z. serrata*; black broken line, *C. camphora*; gray line, plagiophile distribution; gray broken line, planophile distribution; gray dotted line, spherical distribution; and the vertical line and vertical broken line indicate the mean incident zenith angle of the laser beams from the flight track for *Z. serrata* and *C. camphora*, respectively.

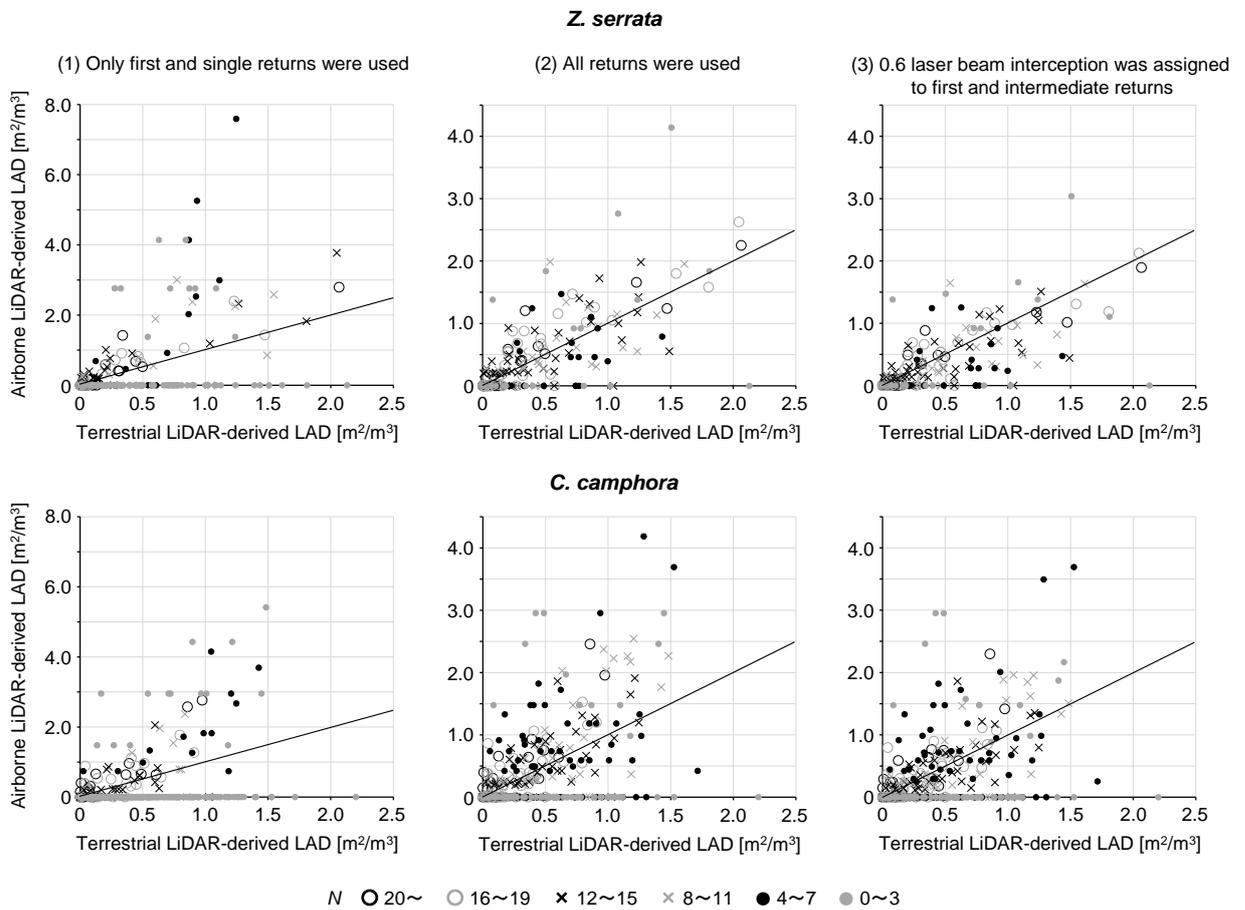

Fig. 7. Scatter plot of airborne LiDAR-derived LAD and terrestrial LiDAR-derived LAD. Results are plotted for each number of incident laser beams on a voxel (*N*). The symbol indicates *N*. Black ring: >20, Gray ring:16–19, Black cross:12–15, Gray cross:8–11, Black circle:4–7, and Gray circle:0–3. Leftmost: Results when only first and single returns were used; Middle: Results when all returns were used; Rightmost: Results when 0.6 laser beam interception was assigned to first and intermediate returns. Top: *Zelkova serrata*, Bottom: *Cinnamomum. camphora*.

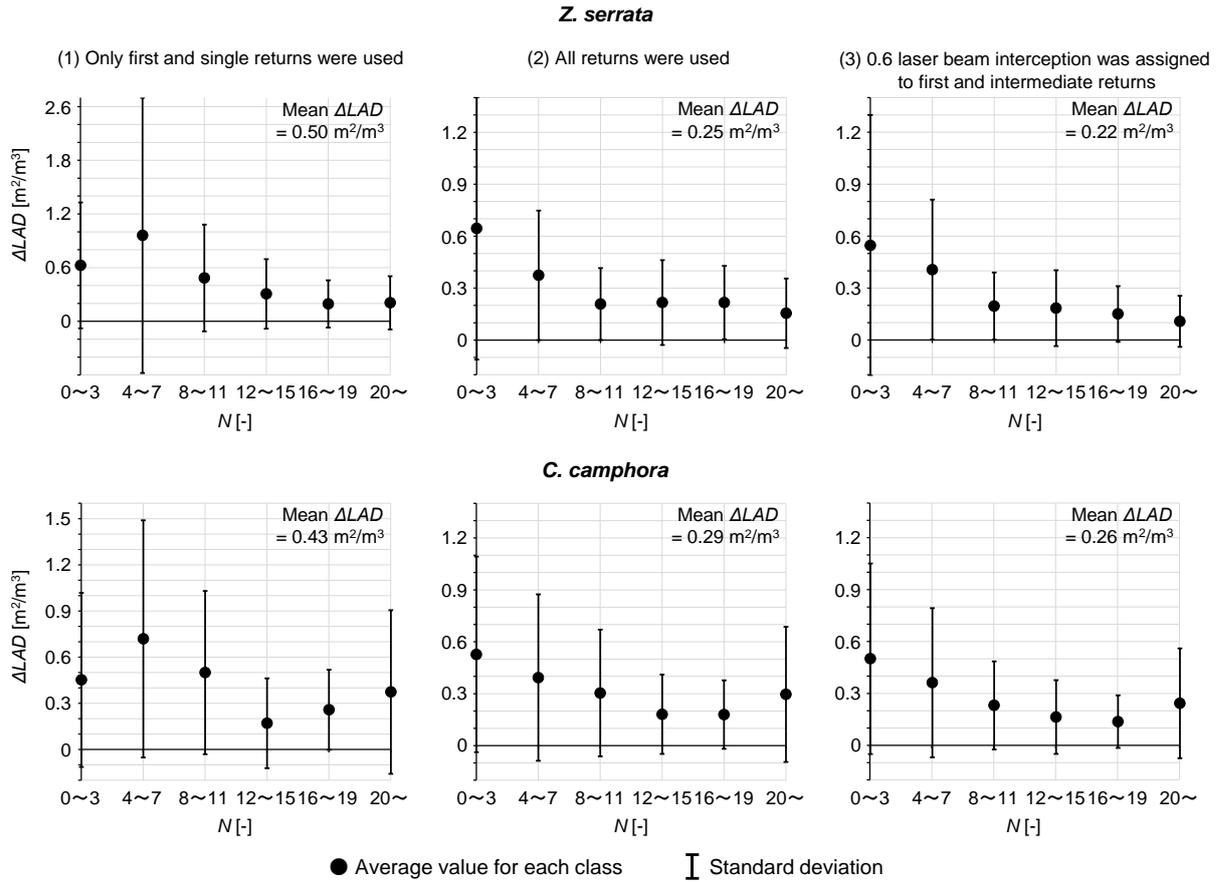

Fig. 8. Variation of absolute difference between airborne LiDAR-derived LAD and terrestrial LiDAR-derived LAD ($\Delta LAD$) according to number of incident laser beams on a voxel ($N$). $\Delta LAD$ was averaged for each $N$ class. Mean $\Delta LAD$ shown at above right of each graph is the mean $\Delta LAD$ for all leaf voxels. This was calculated using the histogram of $N$ for the leaf voxels. Top: *Zelkova. serrata*, Bottom: *Cinnamomum camphora*.

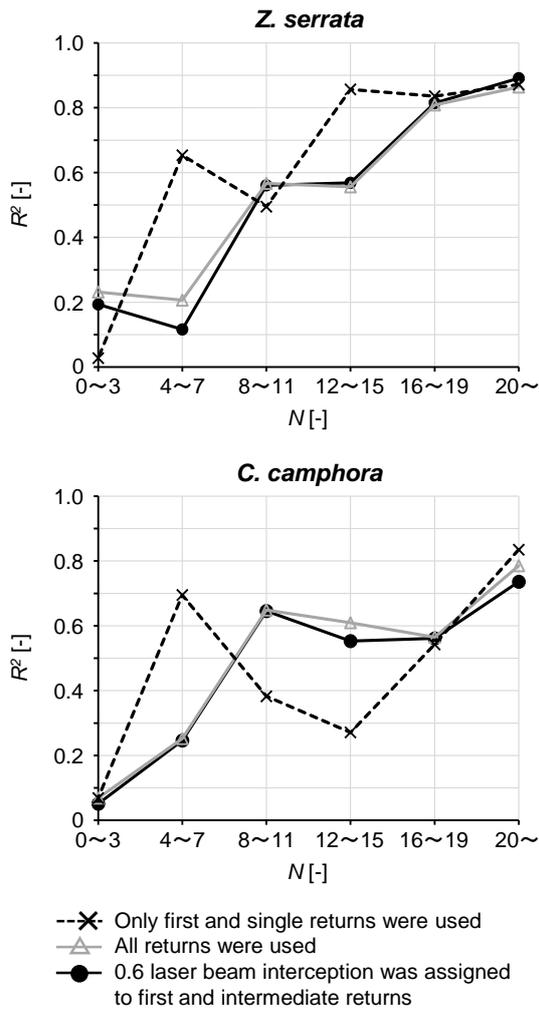

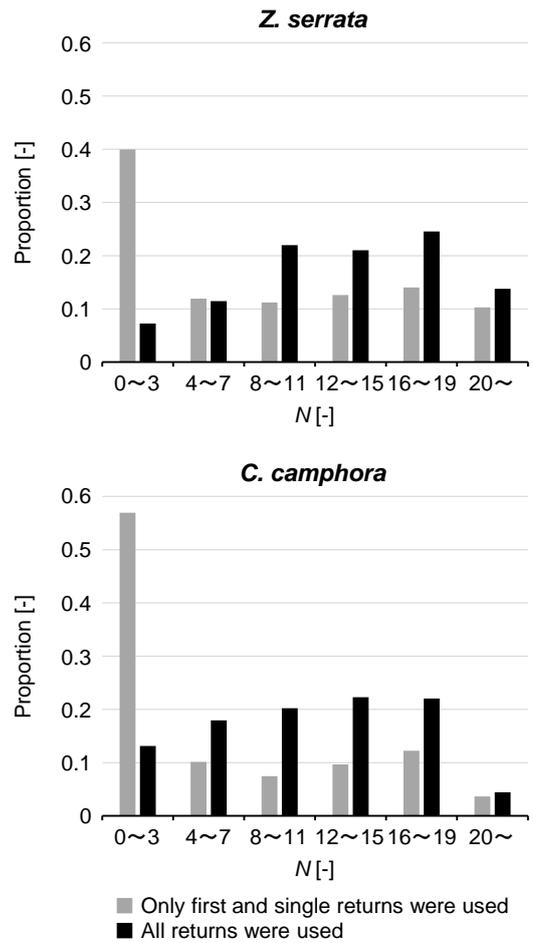

Fig. 9. Variation of $R^2$ according to number of incident laser beams on a voxel ($N$). Top: *Zelkova. serrata*, Bottom: *Cinnamomum camphora*.

Fig. 10. Proportion of number of incident laser beams on a voxel ($N$) for all leaf voxel. Gray: Only first and single returns were used, Black: All returns were used. Top: *Zelkova. serrata*, Bottom: *Cinnamomum camphora*.

Table 1    Airborne LiDAR observation specifications

| | |
|---|---|
| **Date** | September 6, 2010 |
| **Observation system** | SAKURA |
| | (Heliborne system, Nakanihon Air Service) |
| **Altitude** | 350 m |
| **Point spacing on the ground** | 0.25 m (Scan direction), |
| **under the flight track** | 0.2 m (Flight direction) |
| **Scanner** | LMS-Q560 (RIEGL) |
| **Wave length** | 1550 nm |
| **Laser beam divergence** | 0.5 mrad |
| **Ranging accuracy** | 20 mm |
| **Range resolution** | 0.5 m |
| **Number of targets per pulse** | unlimited |

Table 2    Terrestrial LiDAR observation specifications

| | |
|---|---|
| **Date** | September 2, 2010 |
| **Scanner** | VZ-400 (RIEGL) |
| **Scan angle range** | -40–60° (vertical), 360° (horizontal) |
| **Interval between consecutive** | 0.04° |
| **laser shots** | |
| **Point spacing** | 7 mm (vertical plane at 10 m ahead) |
| **Wave length** | 1550 nm |
| **Ranging accuracy** | 5 mm |
| **Laser beam divergence** | 0.3 mrad |